\documentclass[11pt]{article}
\usepackage{epsfig}
\usepackage{amssymb}
\usepackage{psfrag}
\usepackage{xcolor}

\begin{document}

\title{Neutron--neutron scattering length from the reaction $\gamma d \to \pi^+ nn$ \\
 employing chiral perturbation theory}
\author{V.~Lensky$^a$, V.~Baru$^b$, E.~Epelbaum$^{a, c}$, C.~Hanhart$^a$, J.~Haidenbauer$^a$,
\\ A.~Kudryavtsev$^b$, Ulf-G.~Mei\ss ner$^{c, a}$
\vspace{0.5cm}\\
{\small $^a$ Institut f\"{u}r Kernphysik, Forschungszentrum J\"{u}lich GmbH,}\\ 
{\small D--52425 J\"{u}lich, Germany} \\
{\small $^b$ Institute of Theoretical and Experimental Physics,} \\
{\small 117259, B. Cheremushkinskaya 25, Moscow, Russia} \\
{\small $^c$ Helmholtz-Institut f\"{u}r Strahlen- und Kernphysik (Theorie), } \\ 
{\small Universit\"at Bonn, Nu{\ss}allee 14-16, D--53115 Bonn, Germany }
}
\maketitle
\begin{abstract}We discuss the possibility to extract the
  neutron-neutron scattering length $a_{nn}$ from experimental spectra
 on the reaction $\gamma d\to\pi^+ nn$.  The transition operator is
 calculated to high accuracy from chiral perturbation theory. 
 We
 argue that for properly chosen kinematics, the theoretical
 uncertainty of the method can be as low as $0.1$~fm.
\end{abstract}

\vspace{-12.5cm}
\hfill{\tiny FZJ-IKP-TH-2007-13, HISKP-TH-07/12}
\vspace{12.5cm}

\section{Introduction}
\label{intro}
A precise knowledge of the neutron-neutron scattering length $a_{nn}$ is, e.g.,
important for an understanding of the effects of charge symmetry breaking in
nucleon--nucleon forces \cite{Miller}. The scattering length $a_{nn}$
characterizes scattering at low energies. It is related to the on--shell $^1S_0$
scattering amplitude $f^\mathrm{on}$ as
\begin{equation}
f^\mathrm{ on}(p_r)
 = \frac1{p_r\cot \delta(p_r)-ip_r} =
\frac{1}{-a_{nn}^{-1}+\frac{1}{2}\, r_{nn}p_r^2+{\cal O}(p_r^4)-i\,p_r},
\label{effrange}
\end{equation}
where $p_r$ is the relative momentum between the two neutrons, $\delta(p_r)$ the
scattering phase shift in the $^1S_0$ partial wave  and
$r_{nn}$ is the effective range. At low energies the terms of order $p_r^4$ can
be neglected to very high accuracy.  
Obviously, a direct determination of $a_{nn}$ in a scattering experiment is 
extremely difficult due to
the absence of a free neutron target. For this reason, the 
value for $a_{nn}$ is to be obtained from analyses of reactions
where there are three particles in the final state, {\it e.g.}
$\pi^-d\to\gamma nn$ \cite{gabioud,schori,howell} or $nd\to pnn$ 
\cite{gonzales,huhn,gonzalestrotter}.
There is some spread in the results for $a_{nn}$ obtained by the various 
groups. In particular, two independent analyses of the
reaction $nd\to pnn$ give significantly different values for $a_{nn}$,
namely $a_{nn}=-16.1\pm 0.4$ fm \cite{huhn}
and $a_{nn}=-18.7\pm 0.6$ fm \cite{gonzalestrotter}, whereas the latest 
value obtained from the reaction $\pi^-d\to\gamma nn$ is $a_{nn}=-18.5\pm
0.3$ fm \cite{howell}. At the same time, for the proton-proton
scattering length, which is directly accessible, a very recent analysis 
reports $a_{pp}=-17.3 \pm 0.4$ fm \cite{wiringa} after correcting for
electromagnetic effects. This means that even the sign of $\Delta_a=a_{pp}-a_{nn}$
is not fixed.\footnote{Note, that, in contrast to $a_{pp}$,  $a_{nn}$ is not corrected for electromagnetic
effects. However, since those are only
of the order of  0.3 fm \cite{Miller} they are not
relevant for the sign of $\Delta_a$. But they ought to be taken into account 
for determining charge symmetry breaking effects quantitatively.}
It should be mentioned, however, that state of
the art calculations for the binding energy difference of tritium and ${^3}$He 
suggest that $\Delta_a>0$~\cite{Machleidt:2000vh,Nogga:2002qp}.

In the present work we discuss the possibility to determine
$a_{nn}$ from differential cross sections in the reaction $\gamma
d\to\pi^+ nn$. Specifically, we show that one can extract the value of
$a_{nn}$ reliably by fitting the shape of a properly chosen momentum spectrum. 
In this case the main source of inaccuracies, caused by 
uncertainties in the single--nucleon photoproduction multipole $E_{0+}$, 
is largely suppressed. 
Furthermore there is a suppression of the quasi-free pion
production at specific angles. We show that at these angular
configurations the extraction of $a_{nn}$ can be done with minimal
theoretical uncertainty.

Our investigation is based on the recent work of
Ref.~\cite{ourselves} in which
the transition operator for the reaction $\gamma d\to \pi^+ nn$ was
calculated up to order $\chi^{5/2}$ in chiral perturbation theory
(ChPT) with $\chi=m_\pi/M_N \simeq 1/7$, 
where $m_\pi$ ($M_N$) is the pion (nucleon) mass.
Half-integer powers of $\chi$ in the expansion arise from the
unitarity (two-- and three--body) cuts (see also~\cite{baru}). The results
of Ref.~\cite{ourselves} for the total cross section are in very
good agreement with the experimental data. The only input parameter that
entered the calculation was the leading single--nucleon photoproduction multipole
$E_{0+}$, which was fixed from a fourth-order one-loop calculation of 
Bernard et al. \cite{Bernard}. 
The uncertainty in $E_{0+}$ is the main theoretical error in the calculation
presented in Ref.~\cite{ourselves}. 
Besides this transition operator, in the present study 
we use nucleon--nucleon ($NN$) wave functions constructed likewise in the 
framework of ChPT, namely those of the NNLO interaction of Ref.~\cite{epelbaum}. 
This allows us to estimate the theoretical uncertainty which arises from 
variations in the wave functions. In fact,  
as soon as we include consistently all terms up to order $\chi^{5/2}$, we 
expect the ambiguities due to different wave functions not to be larger
than a $\chi^3$ correction, for only at this order the leading counter term
which absorbs these effects enters. 
This expectation is indeed quantitatively confirmed in the concrete 
calculations.

Since we work within chiral perturbation theory we can estimate
the effect of higher orders in terms of established expansion
parameters together with the standard assumption that additional short
ranged operators, that enter at higher orders, behave in accordance
with the power counting (the so-called naturalness assumption). This
method was also applied in Refs.~\cite{phillips,phillips2}, where the
reaction $\pi^-d\to \gamma nn$ was investigated as a tool to extract
$a_{nn}$. However, to know the effect of higher orders for sure, one has
to calculate them. Therefore, to derive a reliable uncertainty
estimate for the extraction of $a_{nn}$ from the $\gamma d$ reaction,
we use our leading order calculation as baseline result and estimate
the theoretical uncertainty from the effects of the higher orders that
we calculated completely. Based on this, we find a theoretical
uncertainty $\delta a_{nn}\lesssim 0.1\ \mbox{fm}.$ We therefore
argue that the reaction $\gamma d\to\pi^+ nn$ 
appears to be a good tool for the extraction of $a_{nn}$.

 To end this section, we remark that in Ref.~\cite{withachot} a method
was proposed to extract scattering lengths from $\gamma d$ induced
meson production. However, this approach should not be used here,
since the momentum transfer is not sufficiently large to use this
method and with our explicit calculation of the transition operator we
can reach a significantly higher accuracy.

\begin{figure}[t]
\begin{center}
\epsfig{file=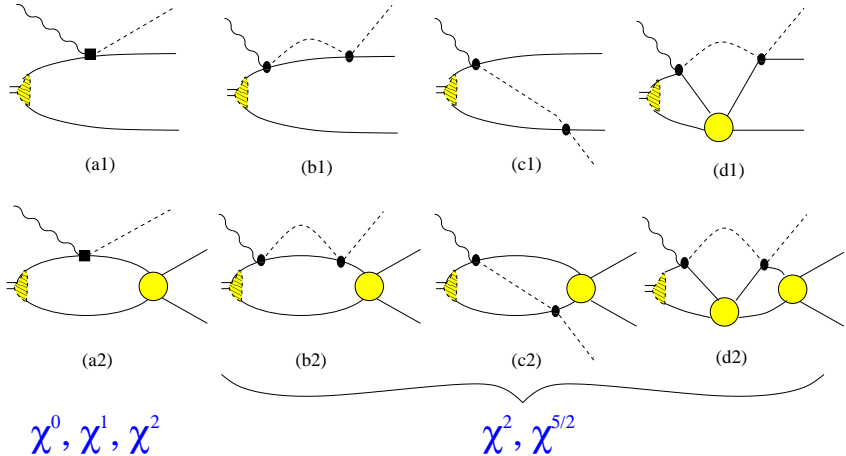, height=6cm, angle=0}
\caption{Diagrams for $\gamma d\to\pi^+nn$. Shown are one--body terms
  $\bigl ($diagram (a) and (b) $\bigr)$, as well as the corresponding
  rescattering contribution (c)---all without and with final state
  interaction. Diagrams (d) shows the class of diagrams with
intermediate $NN$ interaction. Solid, wavy, and dashed
  lines denote nucleons, photons and pions, in order.  Filled squares
  and ellipses stand for the various vertices (see
  Ref.~\cite{ourselves} for the details), the hatched area shows the
  deuteron wave function and the filled circle denotes the $nn$
  scattering amplitude.  Crossed terms (where the external lines are
  interchanged) are not shown explicitly. }
\label{diag}
\end{center}
\end{figure}

\begin{figure}[t]
\begin{center}
\epsfig{file=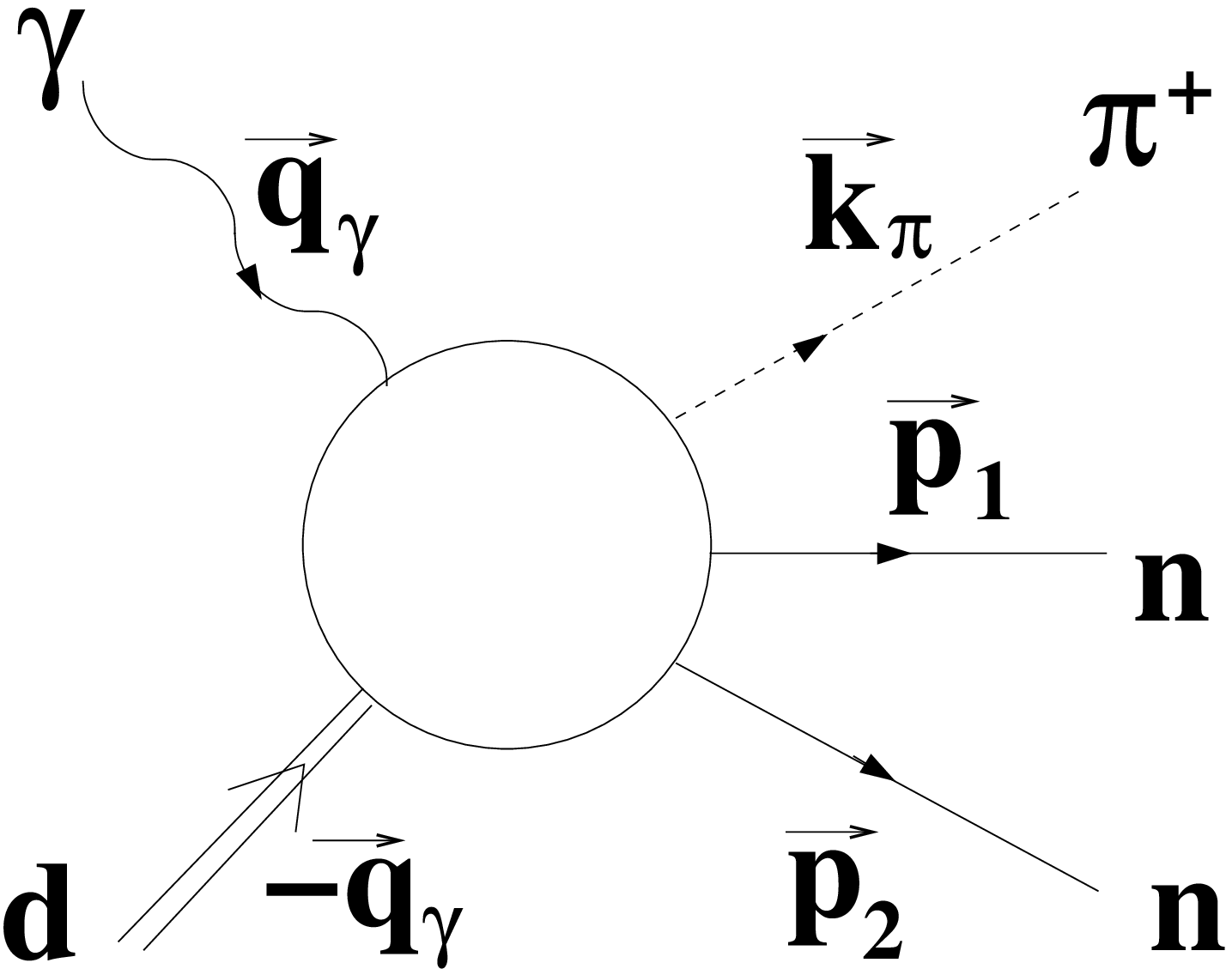, height=4cm, angle=0}
\caption{Kinematical variables for $\gamma d\to\pi^+nn$. The relative
  neutron--neutron momentum is defined as $\vec p_r=\frac{1}{2}(\vec p_1-\vec p_2)$.
}
\label{vec}
\end{center}
\end{figure}

\section{ChPT calculation for {\boldmath$\gamma d\to \pi^+nn$}}\label{chpt}

The diagrams that contribute to the reaction
$\gamma d\to \pi^+nn$ are shown on Fig.~\ref{diag}. The
kinematical variables are defined in  Fig.~\ref{vec}.

Before going into the details some comments are necessary regarding the
relevant scales of the problem. In the near threshold regime of interest here
(excess energies of at most 20 MeV above the pion production threshold) the
outgoing pion momenta are  small compared even  to the pion mass. Thus, in addition
to the conventional expansion parameters of ChPT $m_\pi/\Lambda_\chi$ and
$q_\gamma/\Lambda_\chi$, where $\Lambda_\chi$ denotes the chiral symmetry
breaking scale of order of (and often identified with) the nucleon mass, and
$q_\gamma$ denotes the photon momentum in the center--of--mass system which is of
order of the pion mass, we can also regard $k_\pi/m_\pi$ as small, where
$k_\pi$ denotes the momentum of the outgoing pion.  In what follows we will
perform an expansion in two parameters, namely
$$
\chi_m = m_\pi/M_N \ \mbox{and} \ \chi_Q = k_\pi/m_\pi \ .$$
Obviously, the value of the second parameter depends on the excess energy $Q$. 
The energy regime of interest to us 
corresponds to 
excess energies up to 20 MeV. The maximum value of
$\chi_Q$, $\chi_Q^{max}=\sqrt{2Q/m_\pi}$, at the highest energy considered is thus
about 1/2. Since this is numerically close to $\sqrt{\chi_m}$ we use the
following assignment for the expansion parameter:
\begin{equation}
 \chi \sim \chi_m \sim
\chi_Q^2 \ .
\label{expansion}
\end{equation}The tree level $\gamma p\to \pi^+n$ vertex, as it
appears in diagrams (a1) and (a2) in Fig~\ref{diag} (the vertex is
  labeled as filled square), contributes at leading
  order (order $\chi^0$), and orders $\chi^1$ and $\chi^2$, depending on the
  one--body operator used. Note that the loop diagrams with $\pi N$ rescattering (see diagrams (b),
(c) and (d) in Fig.~\ref{diag}) contribute at order $\chi_m^2$ as well as at 
$\chi_m^2\chi_Q$, $\chi_m^{5/2}$ and at $\chi_m^{1/2}\chi_Q^4$. The origin of the non--integer power of
$\chi$ are the two--body ($\pi N$) and three--body ($\pi NN$)
singularities. Thus, all terms up to $\chi^{5/2}$ are explicitly taken
into account in our calculation of the transition operator.

As 
already emphasized, we employ wave functions
evaluated in the same framework in order to have a fully consistent
calculation. 
In our work, we use the N$^2$LO wave functions corresponding to the 
chiral NN forces introduced in Ref.~\cite{egm} and based on the spectral
function regularization (SFR) scheme \cite{Epelbaum:2003gr}. 
At this order, the NN force receives contributions from one-pion exchange, 
two-pion exchange at the subleading order as well as from all possible
short-range contact interactions with up to two derivatives. 
In addition, the
dominant isospin-breaking correction due to the charged-to-neutral pion mass
difference in the one-pion exchange potential together with the two leading
isospin-breaking S-wave contact interactions were taken into account \cite{egm}. 
The two corresponding low-energy constants were adjusted to reproduce the
scattering lengths $a_{nn}$ and $a_{pp}$. The SFR cutoff $\tilde \Lambda$ is
varied in the range $500 \ldots 700$ MeV. It was argued in
Ref.~\cite{Epelbaum:2003gr} that such a choice for $\tilde \Lambda$ 
provides a natural separation of the long- and short-range parts of the
nuclear force and allows to improve the convergence of the chiral expansion
\cite{Epelbaum:2003gr}. The cutoff $\Lambda$ in the Lippmann-Schwinger
equation is varied in the range $450 \ldots 600$ MeV. For an extensive 
discussion on the choice of $\Lambda$ and $\tilde \Lambda$ the reader is
referred to \cite{epelbaum,egm}.


\section{Differential cross sections: relevant features}
\label{diffxs}

In this section we outline the features of the differential
cross section for unpolarized particles that are important for our considerations.
For later convenience
let us consider the function $F$ proportional to the square of the matrix element as 
well as the five--fold differential cross section 
\begin{equation}
\displaystyle F(p_r,\theta_r,\phi_r,\theta_{\pi},\phi_{\pi})=C\, p_r\ k_\pi(p_r)\overline{|\mathcal{M}(p_r,\theta_r,\phi_r,\theta_{\pi},\phi_{\pi})|^2}
\propto\frac{d^5\sigma(p_r,\theta_r,\phi_r,\theta_{\pi},\phi_{\pi})}{d\Omega_{\vec
     p_r}d\Omega_{\vec k_\pi}dp_r^2} \ ,
\label{XS}
\end{equation}
where $\vec p_r$ ($\vec k_\pi$) stands for the relative momentum of the
two final neutrons (momentum of the final pion) in the center--of--mass
frame, $\theta_r,\ \phi_r$
($\theta_{\pi},\ \phi_{\pi}$) for the corresponding polar and azimuthal angles,
respectively, and $\overline{|\mathcal{M}|^2}$ for the squared and
averaged amplitude. In Eq.~(\ref{XS})  $C$ is an irrelevant dimensionful
constant. 
 In what follows we will consider
only shapes of cross sections and therefore the value
of $C$ is not important for our considerations. The
value of $k_\pi$ at given $p_r$ and excess energy $Q$ is fixed by energy conservation:
\begin{equation}
Q=\frac{p_r^2}{M_N}+\frac{k_\pi^2}{4M_N}+\frac{k_\pi^2}{2m_\pi}\, ,
\label{conserv}
\end{equation}
hence we write $k_\pi(p_r)$ in Eq.~(\ref{XS}).

In the following we choose the momentum $\vec q_\gamma$ of the initial
 photon to be along the $z$--axis. Then the cross sections at a
 certain excess energy $Q$ depend on four variables, namely the magnitude of the
 relative momentum of the two final neutrons $p_r$, the polar angles of the
 vectors $\vec p_r$ and $\vec k_\pi$, and the difference between the
 azimuthal angles of those two momenta. Unpolarized cross sections
are invariant under rotations around the beam axis, which makes
the dependence on the missing angle trivial.

\begin{figure}[t]
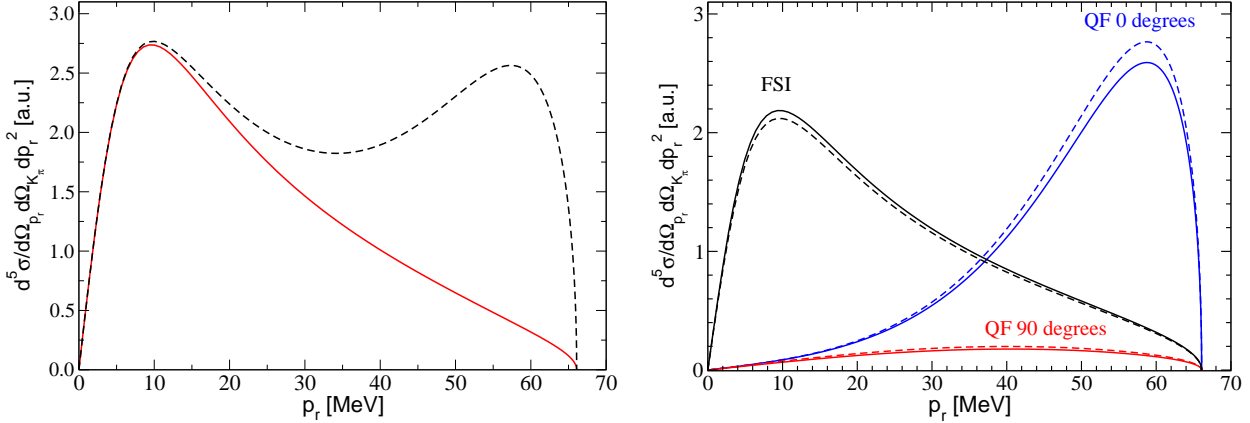

\begin{center}
\begin{tabular}{cc}
\epsfig{file=Q05XS, height=0.33\textwidth,
  angle=0}&\epsfig{file=Q05FSI_QF1, height=0.33\textwidth, angle=0}
\label{difxs}
\end{tabular}
\end{center}
\caption{Left panel: Differential cross section. The solid line corresponds to the 
  configuration when the quasi--free peak is suppressed ($\theta_r=90^\circ$),
   whereas the dashed line corresponds to one of the
  configurations when the quasi--free production amplitude is maximal
  ($\theta_r=0^\circ$). The values of the remaining angles are
  $\theta_{\pi}=135^\circ$, $\phi_r=\phi_{\pi}=0^\circ$; they are the same for
  both curves. Right panel: Differential cross section --- relative strength of QF and FSI
  peaks. Here the dashed curves correspond to the calculation at LO, the 
  solid ones to the calculation at $\chi^{5/2}$. Curves denoted by
  ''FSI'' (''QF'') are obtained by retaining only those diagrams of
  Fig.~\ref{diag} that contain
  (do not contain) the final or the  intermediate nucleon--nucleon
  interaction. The labels ''$0$ degrees'' and ''$90$ degrees'' denote the
  corresponding values of $\theta_r$ for the ''QF'' curves whereas the ''FSI'' curves 
are almost insensitive to this angle. The values or the remaining
  angles are as on the left panel of this figure. The overall
  scale is arbitrary in both panels but the relative
  normalization is the same for all curves.}
\end{figure}

Typical differential cross sections $F$ are shown in Fig.~\ref{difxs}
as a function of $p_r$   at    some   fixed   set    of   angles
$\{\phi_r,\theta_{\pi},\phi_{\pi}\}$  and $Q=5$ MeV for two different
values of $\theta_r$.  One can
see from  this figure that for  the differential cross  section $F$ of
Eq.~(\ref{XS}) there  are two characteristic  regions:
\begin{enumerate} 
\item The region of quasi-free production (QF) at large $p_r$, 
which corresponds to the dominance of those diagrams of Fig.~\ref{diag}
that do not contain the $NN$ interaction in the final or intermediate
states.  In the Appendix we give explicit expressions for the
diagram $a1$ -- the most significant diagram of this type. At large
$p_r$ the pion momentum $k_{\pi}$ is small (see Eq.~(\ref{conserv}))
and the arguments of the deuteron wave function in
Eqs.~(\ref{qfs}) and (\ref{qft}) may become small for particular combinations
of $\pm\: \vec p_r$ and
$\vec q_\gamma/2$.  This feature gives rise to a peak in
the differential cross section at large $p_r$.
\item The region with prominence of the strong $nn$ final--state interaction (FSI) at
 small $p_r$ (in fact, we would have the strongest final state
interaction at zero relative momentum, however the cross section goes
to zero at $p_r=0$ due to the phase space, therefore we see a peak
shape). 
\end{enumerate} 
One can see from Fig.~\ref{difxs} that the FSI peak
depends on the value of $\theta_r$ only marginally, whereas the
quasi-free peak shows significant dependence on this angle. In
particular, the quasi-free production is largely suppressed at
$\theta_r=90^\circ$ --- at this angle the arguments of the wave
functions in both terms in the r.h.s. of Eqs.~(\ref{qfs}) and (\ref{qft})
are large. It can also be seen from Fig.~\ref{difxs} (right panel)
that the effect of higher orders is more important for the quasi-free
production amplitude --- the influence of higher-order effects on the
FSI production is quite small. Another interesting observation is that
the contributions of higher orders change the relative height of the
two peaks -- the FSI peak goes up whereas the QF peak goes down when we 
proceed from the LO calculation to the order $\chi^{5/2}$. In order to 
suppress the distortions of the spectrum due to higher orders in the chiral
expansion, which is the condition for an extraction of $a_{nn}$ with
small theoretical uncertainty, configurations should be chosen where
$\theta_r=90^\circ$.

\begin{figure}[t]
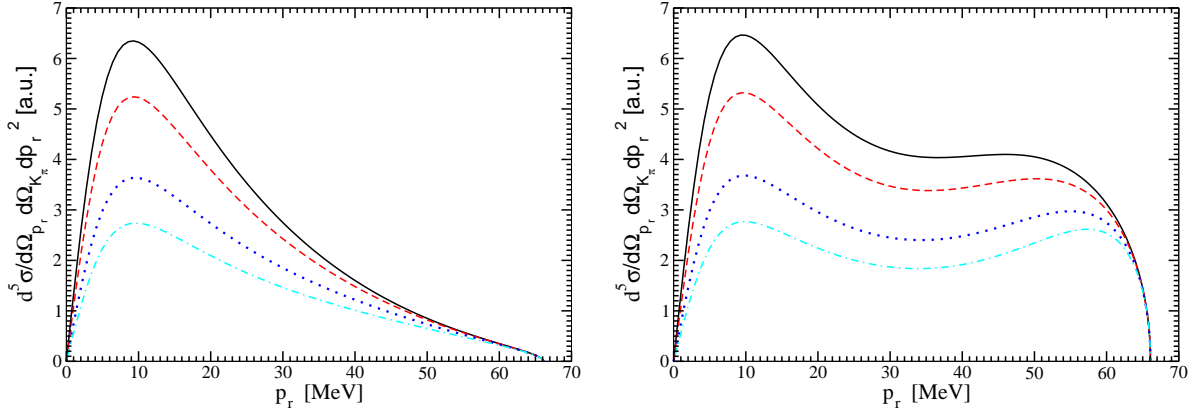

\begin{center}
\begin{tabular}{cc}
\epsfig{file=Q05_NNLO_THETA1_DEPQF, width=0.45\textwidth, angle=0}&\epsfig{file=Q05_NNLO_THETA1_DEPFSI, width=0.45\textwidth, angle=0}

\end{tabular}
\end{center}
\caption{Dependence of the differential cross section on
  $\theta_{\pi}$. The left panel corresponds to the suppressed quasi--free
  amplitude  ($\theta_r=90^\circ$), the right panel to the maximal
  quasi--free amplitude ($\theta_r=0^\circ$). Solid, dashed, dotted,
  and dash-dotted lines correspond to $\theta_{\pi}=0^\circ,\ 45^\circ,\
  90^\circ,\ 135^\circ$ respectively. The values of the remaining angles
are $\phi_r=\phi_{\pi}=0^\circ$. The overall
  scale is arbitrary in both panels but the relative
  normalization is the same for all curves.}\label{theta1dep}
\end{figure}

We now briefly discuss the dependence of the cross section on
the remaining angles  $\theta_{\pi},\phi_{\pi}$ (we always may choose $\phi_r$ to be
zero). The dependence on $\theta_{\pi}$ is illustrated in
Fig.~\ref{theta1dep}. One can see from this figure that the dependence
on $\theta_{\pi}$ is significant for both the quasi--free as
well as the FSI peak.
This can be easily understood from the explicit expressions for the 
matrix elements given in the Appendix keeping in mind
that already at $Q=5$ MeV the maximal value of $k_\pi$ is about
$m_\pi/3$ while $q_\gamma\approx m_\pi$. 
Thus, the momentum transfer to the nucleon pair, $|\vec q_\gamma - \vec k_\pi|$, 
varies in the range $2m_\pi /3$ to $4m_\pi /3$ depending on 
$\theta_\pi$.
Since the $S$-wave deuteron wave function is large only for very small 
arguments, the influence of the direction of
$\vec k_\pi$ is significant.
In addition,  from Fig.~\ref{theta1dep} it follows that a variation of $\theta_{\pi}$
not only changes the magnitude but also the shape of the cross section,
even in the FSI region. This has to be taken into account in the
analysis of any experiment.

In contrast to the polar angles, the dependence of $F$ on $\phi_{\pi}$
is negligible for all configurations (there is no dependence at all
for $\theta_r=0^\circ$ and at $\theta_r=90^\circ$, only the anyway small
QF contribution changes by just 5 \%).

\section{Extraction of \boldmath{$a_{nn}$} and estimate of the
theoretical uncertainty}
\label{extract}

In this section we discuss how to extract the scattering
length from future data on $\gamma d\to \pi^+nn$ as well as
the resulting theoretical uncertainty. Our focus is especially
the latter point. As in the previous section we will only discuss
results at excess energy $Q=5$ MeV. However, the analysis can be repeated analogously
at any excess energy within the range of applicability of the
formalism, i.e. $Q\le 20$ MeV.

\begin{figure}[t]
\begin{center}
\epsfig{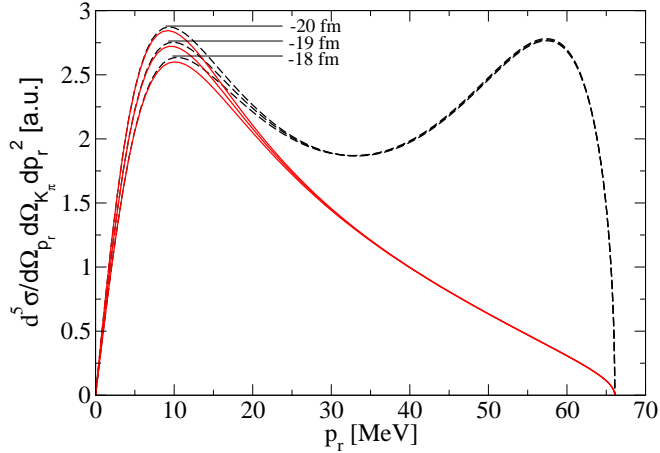}
\caption{The effect of varying the value of $a_{nn}$ on the differential
  cross section. The solid and dashed lines correspond to the same
  angular configurations as in Fig.~\ref{difxs}, left panel. The different values of
  $a_{nn}$ are shown on the figure. The overall
  scale is arbitrary  but the relative
  normalization is the same for all curves.}
\label{ann}
\end{center}
\end{figure}

We are interested in extracting the value of $a_{nn}$, which, in turn,
is a low-energy characteristic of neutron-neutron scattering and
manifests itself in the momentum dependence of the cross section at small values
of the momentum $p_r$. The influence of the value of $a_{nn}$ on the cross section
is illustrated in Fig.~\ref{ann}, where the cross sections are shown for three
different values of $a_{nn}$, namely $-18,\ -19,\ -20\,$fm.
For each value there are two curves, the dashed one corresponds to
 $\theta_r=0^\circ$, and
the solid one to $\theta_r=90^\circ$. One can see from Fig.~\ref{ann} that
the influence of different values of $a_{nn}$ is significant in the
FSI peak and marginal in the quasi--free peak, as one would have
expected.

In the previous section we have shown 
(see right panel in  Fig.~\ref{difxs})  that the relative height of
the quasi-free and the FSI peak changes if the  effects of
higher orders are included in the cross sections. Therefore
those angular configurations are to be preferred, where the
quasi-free production is suppressed.

\begin{figure}[t]
\begin{center}
\epsfig{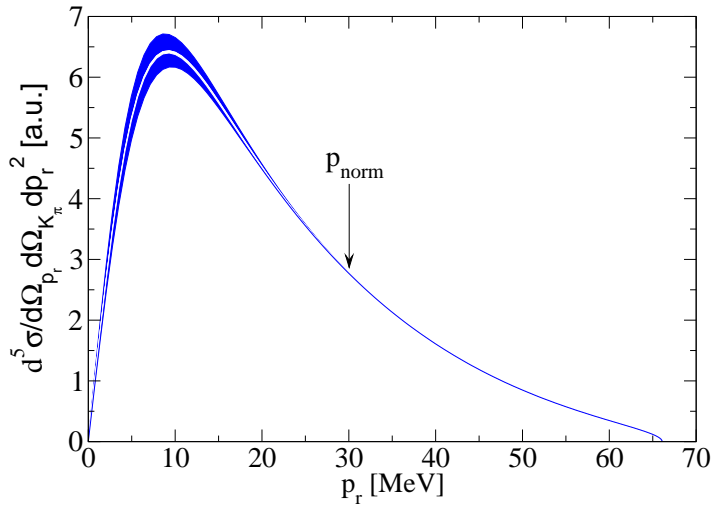}
\end{center}
\caption{ The light (white) band 
is the error band, and dark (blue) band correspond to $\pm 1\ \mathrm{fm}$
shift in the scattering length from the central value $-18.9\
\mathrm{fm}$.}
\label{bands}
\end{figure}

The central point of this study is to demonstrate that there is a large 
sensitivity of the momentum spectra to the scattering length and that
this scattering length can be extracted with 
a small and controlled theoretical uncertainty. 
As outlined in the Introduction, we can estimate this uncertainty 
reliably, because the effect of the higher orders up to $\chi^{5/2}$ 
are calculated completely. 
In order to demonstrate the effect of those higher orders on the
shape of the momentum distribution, in Fig.~\ref{bands} we
show as the light band the spread in the results for
the calculation from LO to $\chi^{5/2}$. The results also include
higher partial waves for the pion as well
as the final $nn$ system. 

There is some sensitivity to the behavior of 
the deuteron wave function at short distances. 
For the reaction $\pi^-d\to \gamma nn$ this sensitivity was
identified as the largest effect at 
N$^3$LO in Ref.~\cite{phillips} \footnote{Within the framework of 
ChPT with a consistent power counting scheme, the quantitative 
impact of the wave-function dependence is governed by the order 
at which a counter term appears that can absorb this model dependence. 
The corresponding counter term for the $\gamma d$ as 
well as the $\pi d$ reaction arises at N$^3$LO.}. 
Guided by that observation
 we include in the uncertainty estimate also the spread
in the results due to the use of different wave functions.
In order to remove the effect of the change in normalization
when, e.g., changing the chiral order, all curves are normalized
at $p_\mathrm{norm}=30$ MeV in Fig.~\ref{bands}.
In the same Figure (with the same normalization) we also show 
the change in the shape that comes from different values of the 
scattering length: the dark band is generated by a variation of
$a_{nn}$ by $\pm 1$ fm around the central value of $-18.9\,$fm.
Clearly, the theoretical uncertainty is negligibly small compared
to the signal of interest.

\begin{figure}[ht]
\begin{center}
\epsfig{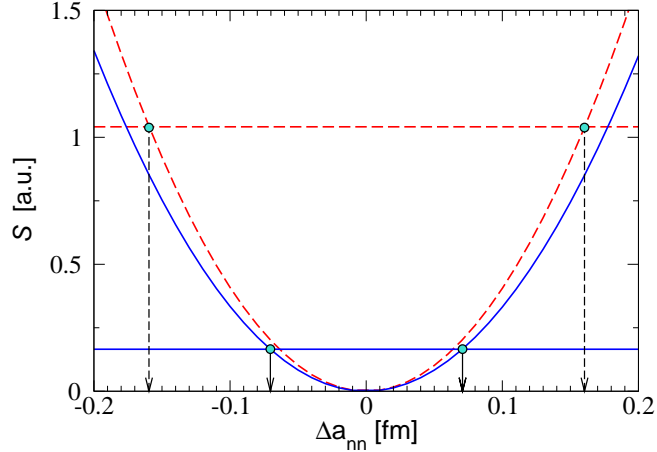}
\caption{The functions ${\mathcal S}(a_{nn}^{(0)},\Phi_\mathrm{max})$ and 
${\mathcal S}(a_{nn}^{(0)}+\Delta a_{nn},\Phi^{(0)})$ are shown by the horizontal and parabolic
curves, respectively. The solid curves are obtained by adding  the weight factor in Eq.~(\ref{funS}) 
that cuts all momenta above 30 MeV  in distinction from the dashed ones.
The calculation is performed for the
scattering length $a_{nn}^{(0)}=-18.9\ \mathrm{fm}$, $\theta_r=90^\circ$, and
$\theta_{\pi}=0^\circ$.  The value of $\Delta a_{nn}$ corresponding to
the crossing point of the horizontal and parabolic curves determines the theoretical
uncertainty of the calculation. }
\label{uncert}
\end{center}
\end{figure}

One way to quantify the theoretical uncertainty is through the 
use of the function ${\mathcal S}$, defined as
\begin{equation}
{\mathcal S}(a_{nn},\Phi)
=\int\limits^{p_\mathrm{max}}_0 dp_r\: \Bigl(F(p_r|a_{nn}^{(0)},\Phi^{(0)})
-N(a_{nn},\Phi)\
F(p_r|a_{nn},\Phi)\Bigr)^2 w(p_r) \, ,
\label{funS}
\end{equation}
where $p_\mathrm{max}=\sqrt{M_N Q}$ is the maximum value of $p_r$,
$F(p_r|a_{nn},\Phi)$ is proportional to the five-fold differential cross 
section as defined in 
Eq.~(\ref{XS}). In the latter we refrained from showing the
angular dependence in favor of the parametric dependence of
the cross section on the $nn$ scattering length $a_{nn}$ as
well as the multi--index $\Phi$, which symbolizes
the dependence of the cross section on the chosen chiral order
and the wave functions used, as outlined above. 
The weight function $w(p_r)$ was introduced to allow
us to suppress particular regions of momenta in the analysis --- the
role of $w(p_r)$ will be discussed in detail below.
For simplicity we may assume that $\mathcal S$ is dimensionless; all
dimensions can be absorbed into the constant $C$ defined
in Eq.~(\ref{XS}).

The value $a_{nn}^{(0)}$  denotes the central value of the scattering length ($-18.9$ fm) for
which we perform the estimate of the theoretical uncertainty\footnote{Note that the theoretical uncertainty 
practically does not change when the central value of the scattering length varies in the relevant interval 
$\pm 1$ fm.} 
 whereas $\Phi^{(0)}$ corresponds to 
the baseline type of calculation, namely leading order with
chiral wave functions as specified in the Appendix.
The relative normalization $N(a_{nn},\Phi)$ is fixed by demanding
that $\mathcal S$ gets minimized for any given pair of parameters 
$a_{nn},\Phi$ ($\partial {\mathcal S}/\partial N=0$). This gives
\begin{equation}
N(a_{nn},\Phi)= \frac{\int\limits^{p_\mathrm{max}}_0
dp_r\: F(p_r|a_{nn}^{(0)},\Phi^{(0)}) F(p_r|a_{nn},\Phi)w(p_r)}
{\int\limits^{p_\mathrm{max}}_0 dp_r\: F^2(p_r|a_{nn},\Phi)w(p_r)} \ .
\label{norm}
\end{equation}
Obviously
$\mathcal S$ is the continuum version of the standard $\chi ^2$
sum, i.e. it characterizes the mean-square deviation from the 
baseline cross section $F(p_r|a_{nn}^{(0)},\Phi^{(0)})$. 
In this way we determine the theoretical uncertainty
in full analogy to the standard method of data analysis.

In order to quantify the theoretical uncertainty we may define
$\Phi_\mathrm{max}$ as that chiral order and choice of wave function,
where ${\mathcal S}(a_{nn}^{(0)},\Phi_\mathrm{max})$ gets maximal:
\begin{equation}
{\mathcal S}(a_{nn}^{(0)},\Phi_\mathrm{max})= \max_\Phi \left\{
{\mathcal S}(a_{nn}^{(0)},\Phi)\right\} \ .
\end{equation}
Therefore ${\mathcal S}(a_{nn}^{(0)},\Phi_\mathrm{max})$ provides an
integral measure of the theoretical uncertainty of the differential cross section.
Demanding that the effect of a change in the scattering length by the amount  $\Delta a_{nn}$
matches that by the inclusion of higher orders {\it etc.}, 
we can identify $\Delta a_{nn}$ as an uncertainty 
in the scattering length.
Expressed in terms of $\mathcal S$, we may define  $\Delta a_{nn}$ via
\begin{equation}
{\mathcal S}(a_{nn}^{(0)}+\Delta a_{nn},\Phi^{(0)})
={\mathcal S}(a_{nn}^{(0)},\Phi_\mathrm{max}) \ .
\end{equation}
 This relation is illustrated in Fig.~\ref{uncert}. The dashed horizontal line
corresponds to ${\mathcal S}(a_{nn}^{(0)},\Phi_\mathrm{max})$, where we
use $w(p_r)=1$.  The dashed parabolic 
line shows the corresponding ${\mathcal S}(a_{nn}^{(0)}+\Delta a_{nn},\Phi^{(0)})$ as a
function of $\Delta a_{nn}$.  The calculation is performed for
$\theta_r=90^\circ$, and $\theta_{\pi}=0^\circ$. The crossing point of the
curves corresponds to $\Delta a_{nn}=0.16\ \mathrm{fm}$, which can be
identified as the theoretical uncertainty for the extraction of the
scattering length. 

In the previous section we showed that the signal region is
located at momenta lower than $30$~MeV. On the other hand, the
theoretical uncertainty of the differential cross section is largest for large values
of $p_r$ due to  the onset of the quasi--free
contribution. In view of these two facts it seems reasonable to use 
such weight functions $w(p_r)$ that suppress the contribution of large momenta.
For instance, we may use $w(p_r)=\Theta(p^\mathrm{ cut}-p_r)$ for the weight function. If we
choose, e.g., $p^\mathrm{ cut}=30$ MeV the theoretical uncertainty of the
extraction of the scattering length reduces to $0.07$ fm, as is demonstrated by the 
solid lines in Fig.~\ref{uncert}. This figure nicely illustrates that the parabolic 
curve that represents the signal changes only very little
when a restriction to small values of $p_r$ is applied. 
At the same time this procedure significantly reduces the value of the uncertainty 
${\mathcal S}(a_{nn}^{(0)},\Phi_\mathrm{max})$.

\begin{figure}[t]
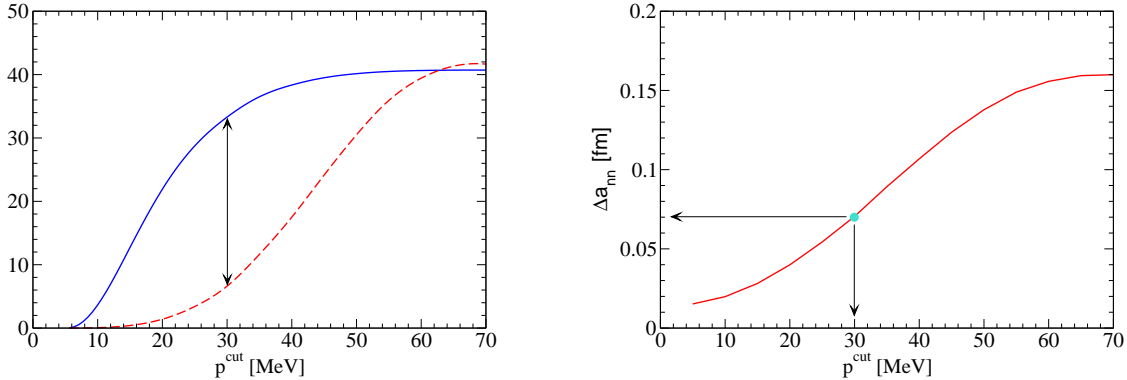

\begin{center}
\epsfig{file=alpha_c_thp000.eps, height=5cm, angle=0} \hspace*{1cm}
\epsfig{file=err_thp000.eps, height=5cm, angle=0}
\caption{Left panel: Comparison of the $p^\mathrm{ cut}$ dependence of functions  ${\mathcal S}(a_{nn}^{(0)},\Phi_\mathrm{max}| p^\mathrm{ cut})$ 
(dashed curve) and $\alpha(p^\mathrm{ cut})$
 (solid curve).  The calculation is performed for the
scattering length $a_{nn}^{(0)}=-18.9\ \mathrm{fm}$, $\theta_r=90^\circ$, and
$\theta_{\pi}=0^\circ$.  
 Right panel: The corresponding theoretical
uncertainty $\Delta a_{nn}$ as a function of $p^\mathrm{ cut}$.
}
\label{uncert2}
\end{center}
\end{figure}

The observation that the dependence of ${\mathcal S}(a_{nn}^{(0)}+\Delta a_{nn},\Phi^{(0)})$
on $\Delta a_{nn}$ is very well approximated by a parabola allows for
a more systematic study of the $p^\mathrm{ cut}$ dependence of the theoretical
uncertainty. We therefore define
\begin{equation}
\alpha(p^\mathrm{ cut})=\frac{{\mathcal S}(a_{nn}^{(0)}+\Delta a_{nn},\Phi^{(0)}|p^\mathrm{ cut})}{(\Delta a_{nn})^2} \ ,
\end{equation}
where the explicit $p^\mathrm{ cut}$ dependence is introduced into the function $\mathcal S$ through
the weight function $w$ as explained above. The dashed and the solid parabola
in Fig.~\ref{uncert} can then be written as $\alpha(p^\mathrm{ cut})\, (\Delta a_{nn})^2$, with
$\alpha (p_\mathrm{max})=41$ fm$^{-2}$ 
 and $\alpha (30~\mathrm{MeV})=33$ fm$^{-2}$. In the left panel of Fig.~\ref{uncert2} we 
show $\alpha(p^\mathrm{ cut})$ as the solid line. 
In the same panel the dashed line represents the measure of the theoretical uncertainty given 
by ${\mathcal S}(a_{nn}^{(0)},\Phi_\mathrm{max}| p^\mathrm{ cut})$,
multiplied by a factor of 40. This figure makes more quantitative the statement
made above: for very small values of $p^\mathrm{ cut}$ we cut into the signal region
and therefore $\alpha$ shows a very rapid variation. However, as soon as 
 $p^\mathrm{ cut}$ is larger than $30$ MeV it goes to a plateau (in the figure indicated by the arrow). On the other
hand, the theoretical uncertainty is monotonously growing once  $p^\mathrm{ cut}$
is larger than $30$ MeV. From this figure we deduce that the ideal
value for  $p^\mathrm{ cut}$ is between $25$ and $40$ MeV. This
translates into a theoretical uncertainty between $0.05$ and $0.1$ fm,
as illustrated in the right panel of the same figure.  
The value of $\theta_{\pi}$ also has some impact on the theoretical
uncertainty, however, in its whole parameter range the estimated
uncertainty stays below $0.1$ fm for  $p^\mathrm{ cut}=30$ MeV.

Clearly, also the experimental data, once they exist, should
be analyzed using a procedure analogous to the one given above.
This means that the scattering length is to be extracted
from a $\chi^2$ fit of the theoretical curves to the data. 
In this work we used the calculation at LO as baseline
result and the results at higher orders to estimate the
theoretical uncertainty. Consequently, we propose to use the momentum spectrum calculated 
at LO in the fitting procedure of the experiment. The corresponding
analytical expressions are given in the Appendix. 
The only parameter
to be adjusted besides the scattering length is the overall
normalization. In this fitting procedure only those
data points should be included that are below a given
$p^\mathrm{ cut}$, in order to keep the theoretical uncertainty small.

\section{Discussion and conclusions}
\label{remarks}
In the previous section it was shown that for the angular
configurations that suppress the quasi-free production the inclusion
of higher order effects (NLO, N$^2$LO, and $\chi^{5/2}$) as well
as the use of different wave functions 
leads only to a
minor change in the momentum dependence of the five-fold differential
cross sections.

Based on this observation we propose to use the momentum spectrum
calculated at LO for the extraction of the neutron--neutron scattering
length from the data. This procedure has the advantage
that the corresponding matrix elements can be given in an analytic
form (see Appendix) that could be used directly
in the Monte Carlo codes for the experiment analysis. In this way
the non--trivial dependence of the spectra on $\theta_\pi$, discussed
above, can be easily controlled. The scattering length can then be extracted
by a two parameter fit to the data where, simultaneously to a variation in
$a_{nn}$, the normalization constant needs to be adjusted.

Note that the leading order calculation basically agrees to 
the expression given in Ref.~\cite{laget} long ago. However, a systematic
and reliable study of the theoretical uncertainties of the extraction
 was possible only within our full calculation up to order $\chi^{5/2}$ in ChPT. 
In this way we could show that
the reaction $\gamma d\to \pi^+nn$ is very well suited for a determination of the $nn$
scattering length. The theoretical uncertainty of order $0.1$ fm
for the extracted scattering length, estimated in this paper,
is of the same order as that claimed for $\pi^-d\to\gamma nn$ 
\cite{phillips2} and $nd\to pnn$~\cite{howell,gonzalestrotter}.

We discussed in detail  the theoretical uncertainty for a fixed excess energy 
of $Q=5$ MeV only, however, it should be clear that the procedure
can be easily repeated for any energy within the range of applicability
of our approach ($Q \leq 20$ MeV).
For example, we checked that the theoretical uncertainty stays
below $0.1$ fm also at $Q=10$ MeV. Note that the number of events in the
signal region scales roughly with $\sqrt{Q}$, the  phase space available for the pion.
It remains to be seen which energy is the best for the corresponding experiment.

We showed that for a proper choice of both kinematics and weight
function $w$, the theoretical uncertainty for the extraction of the
neutron--neutron scattering length from $\gamma d\to\pi^+nn$ can be as
low as $0.1$ fm. It should be stressed, however, that this error was
evaluated most conservatively -- we use our LO calculation as baseline
result and estimate the theoretical uncertainty from the effects of
the higher orders that we calculated completely.  This error can be
significantly reduced by further studies.  For
example, if we include in the uncertainty estimate only the spread in
the results due to the use of different wave functions, which is
identified as the largest effect at N$^3$LO for the reaction $\pi^-
d\to \gamma nn$~\cite{phillips}, the theoretical uncertainty of the
extracted scattering length reduces by one order of magnitude.  This
indicates that the theoretical uncertainty is indeed under control.
However to put this N$^3$LO estimation on more solid ground a
complete calculation should be performed to this order.
Most of the  
operators that are relevant at this order are the same as those of $\pi^-d\to
\gamma nn$, given explicitly in Ref.~\cite{anders3}. One
counter term enters, which can be fixed from other processes~\cite{phillips2}, e.g., from $nd$
scattering~\cite{andreas}, 
the reaction 
$NN\to NN\pi$~\cite{chiralpwaves}, or
from weak decays~\cite{phillips2}.  Once this is done we may use our calculation to
order $\chi^{5/2}$ as baseline result and estimate the theoretical
uncertainty from the then available N$^3$LO calculation.

Although we have identified the angles $\theta_r=90^{\circ}$ as
the preferred kinematics, also other configurations could be studied
in order to control the systematics. However, then the spectra calculated
at $\chi^{5/2}$ should be used in the analysis.

\section*{Acknowledgments}
We thank A.~Bernstein for useful discussions and interest in this work. We also thank
D.~R.~Phillips and A.~G{\aa}rdestig for helpful discussions.
This research is part of the EU Integrated Infrastructure Initiative
Hadron Physics Project under contract number RII3-CT-2004-506078, and
was supported also by the DFG-RFBR grant no. 05-02-04012 (436 RUS
113/820/0-1(R)) and the DFG SFB/TR 16 "Subnuclear Structure of Matter".  A.~K. and
V.~B. acknowledge the support of the Federal Program of the Russian
Ministry of Industry, Science, and Technology No 02.434.11.7091.
E.~E. acknowledges the support of the Helmholtz Association 
(contract no. VH-NG-222). 

\appendix

\section{Leading amplitudes}
\label{ampl}

\def\theequation{\Alph{section}.\arabic{equation}}
\setcounter{equation}{0}

In this appendix we give explicit expressions for the
amplitudes that appear at leading order in the calculation
for $\gamma d\to\pi^+ nn$. As outlined in the 
main text these expressions can be used directly
in the analysis of the data, once available. In addition,
they should also proof useful  for the design of the
experiment. Note, as outlined in the text, only
near $\theta_r=90^\circ$ the leading order calculation gives
a sufficiently accurate representation of the spectra. At
all other angles one should use the complete calculation.

At leading order only diagrams $a1$ and $a2$ of Fig.~\ref{diag}
contribute. Since only the momentum dependence of the  amplitudes is relevant for the  
 experimental analysis we drop an overall factor  compared to Ref.~\cite{ourselves}.
The corresponding amplitudes read:
\begin{eqnarray}
M_{a1}^{{s}}&=&\left(u(\vec p_r-\vec k_\pi/2+\vec q_\gamma/2)
+u(-\vec p_r-\vec k_\pi/2+\vec q_\gamma/2)\right)\label{qfs}\\
M_{a1}^{{t}}&=&\left(u(\vec p_r-\vec k_\pi/2+\vec q_\gamma/2)
-u(-\vec p_r-\vec k_\pi/2+\vec q_\gamma/2)\right)\label{qft}\\ \nonumber
M_{a2}&=&8\pi\,\frac{f^\mathrm{ on}(p_r)}{g(p_r)}\int\frac{d^3\:p}{(2\pi)^3}
\frac{u(\vec p-\vec k_\pi/2+\vec q_\gamma/2)\; g(p)}{p^2-p_r^2-i0}\, \\
&=& \frac{f^\mathrm{ on}(p_r)}{iq_{\pi\gamma}\: g(p_r)}\sum_{ij}\frac{C_iD_j}{p_r^2+\beta_j^2}
\ \ln\left(\frac{\alpha_i-ip_r+iq_{\pi\gamma}}{\alpha_i-ip_r-iq_{\pi\gamma}}\cdot
\frac{\alpha_i+\beta_j-iq_{\pi\gamma}}{\alpha_i+\beta_j+iq_{\pi\gamma}}\right)
\label{fsi}
\end{eqnarray}
where $u(\vec p)$ denotes the $S$--wave part
of the deuteron wave function in momentum space.
We checked by  explicit calculations that the inclusion of the deuteron $D$-wave 
changes only the absolute scale of the differential cross sections but not its momentum dependence. Thus, the 
$D$-wave contribution is not taken into account in the parameterization. 
The quantity  $q_{\pi\gamma}$ is defined as $q_{\pi\gamma}=|\vec k_{\pi}-\vec q_{\gamma}|/2$.
The labels ${s}$
and ${t}$ stand for spin singlet and triplet final two-nucleon
states, respectively --- we do not write out the corresponding spin structures.
We take into account only the $^1S_0$ partial wave in the final state
interaction.  For a discussion of the effect of $nn$ $P$--waves see
Ref.~\cite{ourselves}.

To derive the expression for $M_{a2}$ we used 
the fact 
that the neutron--neutron scattering amplitude can be represented to high accuracy 
in separable form \cite{ourselves,JH}. The neutron--neutron
scattering amplitude, $f(p,k;E)$, can be written in half off--shell
kinematics as
\begin{equation}
f(p,k;k^2/M_N)= \frac{2\pi^2 M_N g(p)g(k)}{1 - M_N \int d^3 q \frac{g^2(q)}{q^2-k^2-i0}} 
= f^\mathrm{ on}(k) \frac{g(p)}{g(k)} \ , 
\label{fdef}
\end{equation}
where the corresponding on--shell amplitude $f^\mathrm{ on}(k)$ can then be expressed 
in terms of the scattering phase--shifts through
$$
f^\mathrm{ on}(k)=f(k,k;k^2/M_N) = \frac1{k\cot \delta(k)-ik} \ .
$$
For small momenta one can use the effective range expansion for $k\cot \delta=-1/a_{nn}+r_{nn} 
k^2/2 + {\cal O}(k^4)$,
in agreement with Eq.~(\ref{effrange}).
Here  $a_{nn}$ is the parameter to be fitted to the data and 
$r_{nn}=2.76\;$fm. 
We checked that changing the value of $r_{nn}$ within the bounds
allowed ($\pm 0.1$ fm~\cite{Miller}) leads to 
negligible effects on the extraction of the scattering length. 
In this way we expressed the matrix element explicitly
in terms of the scattering length. 
We checked that the ratio $g(p)/g(k)$ in Eq.~(\ref{fdef}) does not change
when we vary the scattering length within acceptable range bounds.

In order to evaluate the convolution of the deuteron wave function
with the $nn$ final state interaction analytically, we needed
to employ the following parameterizations for
the  $^1S_0$ $nn$ form factor $g(p)$ (see Eq.~(\ref{fdef})) and the S-wave deuteron wave function 
\begin{equation}
\nonumber
g(p)=\sum_i\frac{D_i}{p^2+\beta_i^2};\ \ \ \ 
u(p)=\sum_i\frac{C_i}{p^2+\alpha_i^2};
\end{equation}
where the parameters corresponding to the ChPT  calculation at N$^2$LO with cut offs 
\{$\Lambda, \tilde \Lambda\}$  =  $\{550\ $MeV, $\; 600\ $MeV\} (see Ref~\cite{egm} for details) 
are listed in Table \ref{separable}. Note that the coefficients in the parameterization of
the wave function have to fulfill the relation $\sum C_i=0$ in order to ensure 
the regularity of the deuteron wave function 
at the origin in coordinate space \cite{Lacombe}.

\begin{table}[t]
\begin{center}
\begin{tabular}{|r|cc|cc|}
\hline
&\multicolumn{2}{c|}{$^1S_0$ form factor}&\multicolumn{2}{c|}{S-wave deuteron w.f.}\\
\hline
& $\beta_i$ [MeV] & $D_i$ [MeV] & $\alpha_i$ [MeV]& $C_i$ [MeV$^{1/2}$] \\
\hline
1&    164.53278   &    31.101228       &             45.334919    &    43.543212\\    
2&    246.85751   &   -1310.3056       &             242.66091    &   -35.643003\\    
3&    329.18224   &    9455.9603       &             439.98691    &    419.25214\\    
4&    411.50697   &   -9666.0268       &             637.31291    &   -1833.4708\\    
5&    493.83170   &   -55571.615       &             834.63891    &   -3710.8173\\    
6&    576.15643   &    64600.071       &             1031.9649    &    24903.150\\    
7&    658.48116   &    149128.85       &             1229.2909    &   -31673.576\\   
8&    740.80589   &   -84844.967       &             1426.6169    &    26476.636\\   
9&    823.13062   &   -295594.17       &             1623.9429    &   -118733.48\\   
10&   905.45536   &   -30332.710       &             1821.2689    &    259759.15\\   
11&   987.78009   &    560829.89       &             2018.5949    &   -223816.07\\   
12&   1070.1048   &   -307006.25       &             2215.9209    &   $-\sum_{i=1}^{11}C_i$\\    
\hline                                                          
\end{tabular}
\caption{Parameters of the $^1S_0$ form factor and the S-wave deuteron wave function for the separable representation 
of the N$^2$LO chiral $NN$ potential. 
}
\label{separable}
\end{center}
\end{table}

The squared and averaged amplitude to be used in the expression for the differential
cross section, defined in Eq.~(\ref{XS}) is
\begin{equation}
\overline{|\mathcal{M}(p_r,\theta_r,\phi_r,\theta_{\pi},\phi_{\pi})_\mathrm{{fit}}|^2}=
\left|M_{a1}^{{s}}+M_{a2}\right|^2+2\left|M_{a1}^{{t}}\right|^2 \ .
\end{equation}

In a fit to data two parameters are to be adjusted, namely the overall normalization
$C$ of  Eq.~(\ref{XS}) and the object of desire, $a_{nn}$.


\bigskip


\begin{thebibliography}{99}
\bibitem{Miller}
G. Miller, B. Nefkens and I. {\v S}laus. Phys. Rep. {\bf 194}, 1 (1990).
\bibitem{gabioud} B.~Gabioud et al., Nucl. Phys. A {\bf 420}, 496 (1984).
\bibitem{schori} O.~Schori et al., Phys. Rev. C {\bf 35}, 2252 (1987).
\bibitem{howell}C.~R.~Howell  et al., Phys.~Lett.~B {\bf 444},
  252 (1998).
\bibitem{gonzales}D.~E.~Gonz\'ales~Trotter et al.,
Phys.~Rev.~Lett. {\bf 83}, 3788 (1999). 
\bibitem{huhn}V.~Huhn  et al., Phys.~Rev.~Lett. {\bf 85}, 1190
  (2000); Phys.~Rev.~C {\bf 63}, 014003 (2000).
\bibitem{gonzalestrotter}D.~E.~Gonz\'ales~Trotter et al.,
  Phys.~Rev.~C {\bf 73}, 034001 (2006).
\bibitem{wiringa}R.~B.~Wiringa, V.~G.~J.~Stoks and R.~Schiavilla,
  Phys.~Rev.~C {\bf 51}, 38 (1995) [arXiv: nucl-th/9408016].
\bibitem{Machleidt:2000vh}
  R.~Machleidt and H.~M{\"u}ther,
  Phys.\ Rev.\  C {\bf 63}, 034005 (2001)
  [arXiv:nucl-th/0011057].
\bibitem{Nogga:2002qp}A.~Nogga et al., Phys.~Rev.~C {\bf 67}, 034004 (2003) 
  [arXiv:nucl-th/0202037].
\bibitem{ourselves}V.~Lensky, V.~Baru, J.~Haidenbauer, C.~Hanhart,
  A.~Kudryavtsev and U.-G.~Mei\ss ner, Eur.~Phys.~J.~A~{\bf 26}, 107
  (2005)  [arXiv: nucl-th/0505039].
\bibitem{baru}V.~Baru, C.~Hanhart, A.~E.~Kudryavtsev and U.-G.~Mei\ss
  ner, Phys. Lett. B {\bf 589}, 118 (2004)   [arXiv: nucl-th/0402027].
\bibitem{Bernard}   V.~Bernard, N.~Kaiser and U.-G.~Mei\ss ner,
  Phys.\ Lett.\  B {\bf 383}, 116 (1996)  [arXiv:hep-ph/9603278].
\bibitem{epelbaum}E.~Epelbaum, Prog.~Part.~Nucl.~Phys. {\bf 57}, 654 (2006)  
[arXiv: nucl-th/0505032].
\bibitem{phillips}A.~G\aa rdestig and D.~R.~Phillips, Phys.~Rev.~C {\bf
  73}, 014002 (2006)    [arXiv: nucl-th/0501049].
\bibitem{phillips2}
  A.~G\aa rdestig and D.~R.~Phillips,
  Phys.\ Rev.\ Lett.\  {\bf 96} (2006) 232301
  [arXiv:nucl-th/0603045].
\bibitem{withachot}
  A.~Gasparyan, J.~Haidenbauer, C.~Hanhart and K.~Miyagawa,
  arXiv:nucl-th/0701090;  Eur.~Phys.~J.~A in print.
\bibitem{egm} E.~Epelbaum, W.~Gl\"ockle and U.-G.~Mei\ss ner,
  Nucl.~Phys.~A {\bf 747}, 362 (2005) [arXiv: nucl-th/0405048]
\bibitem{Epelbaum:2003gr}
  E.~Epelbaum, W.~Gl\"ockle and U.-G.~Mei{\ss}ner,
  Eur.~Phys.~J.~A {\bf 19}, 125 (2004) 125
  [arXiv:nucl-th/0304037].
\bibitem{laget} J. M. Laget, Phys. Rep. {\bf 69} (1981) 1. 
\bibitem{anders3}
  A.~G\aa rdestig,
  Phys.\ Rev.\  C {\bf 74}, 017001 (2006)
  [arXiv:nucl-th/0604035].
\bibitem{andreas}
 E.~Epelbaum, A.~Nogga, W.~Gl\"ockle, H.~Kamada, U.-G.~Mei\ss ner and H.~Witala,
  Phys.\ Rev.\  C {\bf 66}, 064001 (2002)
  [arXiv:nucl-th/0208023].
\bibitem{chiralpwaves}
 C.~Hanhart, U.~van Kolck and G.~A.~Miller,
  Phys.\ Rev.\ Lett.\  {\bf 85} (2000) 2905
  [arXiv:nucl-th/0004033].
\bibitem{JH} J. Haidenbauer and W. Plessas, 
 Phys.\ Rev.\  C {\bf 30}, 1822 (1984).
\bibitem{Lacombe} 
  M.~Lacombe, B.~Loiseau, R.~Vinh Mau, J.~Cote, P.~Pires and R.~de Tourreil,
  Phys.\ Lett.\  B {\bf 101}, 139 (1981).

\end{thebibliography}
\end{document}